\documentclass{ifacconf}  
\usepackage{natbib}

\usepackage{amsmath} 
\usepackage{amssymb}  

\usepackage{booktabs}
\usepackage{algorithm}
\usepackage{algpseudocode} 
\usepackage{tikz}
\usepackage{mathtools}

\usetikzlibrary{shapes,arrows,calc,fit}
\tikzset{
        block/.style = {draw, rectangle,
            minimum height=1cm,
            minimum width=2cm},
        input/.style = {coordinate,node distance=1cm},
        output/.style = {coordinate,node distance=4cm},
        arrow/.style={draw, -latex,node distance=2cm},
        pinstyle/.style = {pin edge={latex-, black,node distance=2cm}},
        sum/.style = {draw, circle, node distance=1cm},
    }
\usepackage{pgfplots} 
\pgfplotsset{compat=newest} 
\pgfplotsset{plot coordinates/math parser=false} 
\newlength\figureheight%
\newlength\figurewidth%
\pgfplotsset{
    every axis plot post/.style={
        line join=round
    }
}

\usepackage{tikz}
\usepackage{tikzscale}

\definecolor{myorange}{cmyk}{0,0.35,0.85,0} 
\definecolor{mypurple}{cmyk}{0.5,1,0,0} 

\definecolor{matblue1}{rgb}{0,0.4470,0.7410}
\definecolor{matred1}{rgb}{0.85,0.325,0.098}
\definecolor{matyel1}{rgb}{0.9290, 0.6940, 0.1250}
\definecolor{matpur1}{rgb}{0.4940, 0.1840, 0.5560}
\definecolor{matgre1}{rgb}{0.4660, 0.6740, 0.1880}
\definecolor{matblue2}{rgb}{0.3010, 0.7450, 0.9330}
\definecolor{matred2}{rgb}{0.6350, 0.0780, 0.1840}
\definecolor{matgrey1}{rgb}{0.5, 0.6, 0.7}
\definecolor{matpink1}{rgb}{1, 0.07, 0.65}
\definecolor{matblue3}{rgb}{0.07, 0.62, 1}
\definecolor{gray09}{rgb}{0.9, 0.9, 0.9}

    \definecolor{mblue}{rgb}{0,0.447,0.741}
    \definecolor{mred}{rgb}{0.85,0.325,0.098}
    \definecolor{myellow}{rgb}{0.9290,0.6940,0.1250}
    \definecolor{mmagenta}{rgb}{1,0,1}
    \definecolor{mgreen}{rgb}{0.4460,0.6740,0.1880}
    \definecolor{mgrey}{rgb}{0.6,0.6,0.6}
    \definecolor{mpurple}{rgb}{0.4940, 0.1840, 0.5560}

    \definecolor{matbluel}{rgb}{0,0.6732,1}
    \definecolor{matyeld}{rgb}{0.2787,0.2082,0.0375}

    \tikzset{cross/.style={cross out, draw=black, minimum size=2*(#1-\pgflinewidth), inner sep=0pt, outer sep=0pt}, cross/.default={1pt}}
    
    \usetikzlibrary{shapes}

\newcommand{\reddash}{\raisebox{2pt}{\tikz{\draw[-,matred1,dashed,line width = 0.9pt](0,0) -- (3mm,0);}}}

\newcommand{\blackdash}{\raisebox{2pt}{\tikz{\draw[-,black,dashed,line width = 0.9pt](0,0) -- (3mm,0);}}}

\newcommand{\blueline}{\raisebox{2pt}{\tikz{\draw[-,matblue1,solid,line width = 0.9pt](0,0) -- (3mm,0);}}}
\newcommand{\redline}{\raisebox{2pt}{\tikz{\draw[-,matred1,solid,line width = 0.9pt](0,0) -- (3mm,0);}}}

\usepackage{tikz}
\usepackage{xcolor}

\newlength{\boxplotlinewidth} 
    \usepackage{enumitem}

\usepackage{ifthen}
\usepackage{xspace}
\newboolean{thesismode}
\setboolean{thesismode}{false}

\newcommand{\manuscript}{%
    \ifthenelse{\boolean{thesismode}}{chapter\xspace}{paper\xspace}%
}

\usepackage{eso-pic}

\AddToShipoutPictureBG*{%
	\AtPageUpperLeft{%
		\setlength\unitlength{1in}%
		\hspace*{\dimexpr0.5\paperwidth\relax}
		\makebox(0,-2)[c]{\begin{tabular}{c c}
				Max van Meer, Self-Calibrating Position Measurements: Applied to Imperfect Hall Sensors, \\
				Accepted for the joint 10th IFAC Symposium on Mechatronic Systems \& 14th IFAC Symposium on Robotics,\\Uploaded to ArXiv \today
		\end{tabular}}
}}

\usepackage{calc}  

\newcommand{\figfrac}{0.7}
\newlength{\mylinewidth}

\usepackage{xstring} 
\newcommand{\firstword}[1]{%
  \ifthenelse{\boolean{thesismode}}{%
    #1%
  }{%
    \StrLeft{#1}{1}[\firstchar]%
    \StrGobbleLeft{#1}{1}[\restofword]%
    \IEEEPARstart{\firstchar}{\MakeLowercase{\restofword}}%
  }%
}
\begin{document}
\sloppy
\begin{frontmatter}
\title{Self-Calibrating Position Measurements:\\Applied to Imperfect Hall Sensors\thanksref{footnoteinfo}}

\author[TUE]{Max van Meer} 
\author[TUE]{Marijn van Noije} 
\author[TUE]{Koen Tiels} 
\author[Sioux]{Enzo Evers} 
\author[TUE,Sioux]{Lennart Blanken} 
\author[TUE,TNO]{Gert Witvoet} 
\author[TUE,Delft]{Tom Oomen}

\address[TUE]{Control Systems Technology section, Department of Mechanical Engineering, Eindhoven University of Technology, The Netherlands (e-mail: m.v.meer@tue.nl).}
\address[Sioux]{Mechatronics department, Sioux Technologies B.V., Eindhoven, The Netherlands.}
\address[TNO]{Department of Optomechatronics, TNO, Delft, The Netherlands.}
\address[Delft]{Delft Center for Systems and Control, Delft University of Technology, Delft, The Netherlands.}

\thanks[footnoteinfo]{This work is part of the research programme VIDI with project number 15698, which is (partly) financed by the Netherlands Organisation for Scientific Research (NWO). In addition, this research has received funding from the ECSEL Joint Undertaking under grant agreement 101007311 (IMOCO4.E). The Joint Undertaking receives support from the European Union's Horizon 2020 research and innovation program.} 
 
\begin{abstract}
Linear Hall sensors are a cost-effective alternative to optical encoders for measuring the rotor positions of actuators, with the main challenge being that they exhibit position-dependent inaccuracies resulting from manufacturing tolerances. 
This paper develops a data-driven calibration procedure for linear analog Hall sensors that enables accurate online estimates of the rotor angle without requiring expensive external encoders.
The approach combines closed-loop data collection with nonlinear identification to obtain an accurate model of the sensor inaccuracies, which is subsequently used for online compensation. 
Simulation results show that when the flux density model structure is known, measurement errors are reduced to the sensor noise floor, and experiments on an industrial setup demonstrate a factor of 2.6 reduction in the root-mean-square measurement error. 
These results confirm that Hall sensor inaccuracies can be calibrated even when no external encoder is available, improving their practical applicability.
\end{abstract}

\end{frontmatter}

\setlength{\mylinewidth}{\ifthesismode \figfrac\linewidth \else \linewidth \fi}

\section{Introduction}
Accurate position measurements are key in high-performance actuators for applications such as semiconductor manufacturing or optical satellite communication~\citep{Mack2007,Kramer2020}. These actuators must meet strict positioning requirements, often in the micrometer or microradian range, to achieve accurate control performance~\citep{Oomen2018}. 
Meanwhile, the demand for mass-produced solutions has created a need for more economical sensors that still meet these requirements.

\begin{figure}
    \centering
    \includegraphics[width=0.75\mylinewidth]{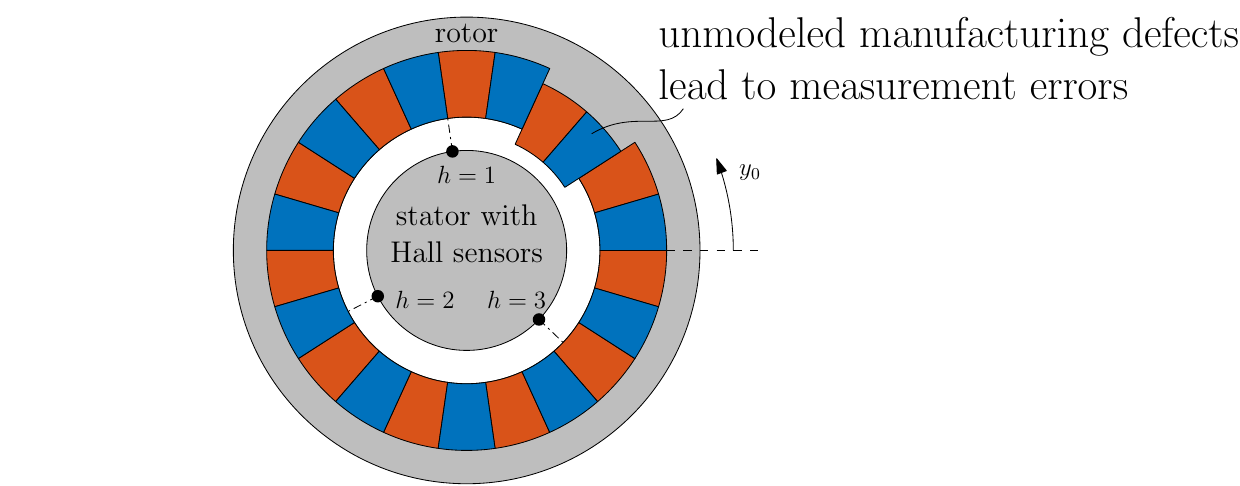}
    \caption{Experimental setup: Linear Hall sensors $h$ on the stator measure flux density $d_h$ from rotor-mounted magnets. Blue and red blocks indicate south and north poles. The flux density depends on the rotor position $y_0$, but reconstructing $y \approx y_0$ is complicated by unmodeled manufacturing defects. Stator windings are omitted from the scheme for simplicity.}
    \label{fig:hall_scheme}
\end{figure}

Figure~\ref*{fig:hall_scheme} depicts a set of Linear Hall sensors on a rotor, which offer a promising alternative to costly, high-resolution encoders for electric actuators. A Hall sensor outputs a voltage proportional to the local magnetic flux density, which can be processed to estimate the rotor angle~\citep{Ramsden2006,Liu2008}. Compared to high-end encoders, linear Hall sensors are cheaper, more compact, and easier to integrate in large volumes.

\begin{figure}
    \centering
    \includegraphics[width=\mylinewidth]{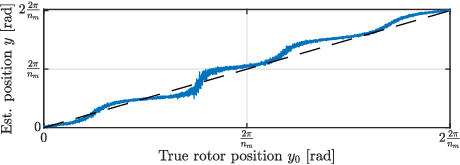}
    \caption{Illustrative example of a position-dependent measurement inaccuracy, plotted along two out of $n_m$ pole-pairs. When the position is reconstructed (\protect\blueline) from flux density signals while neglecting higher order harmonics, the estimate of true rotor angle (\protect\blackdash) is not accurate and potentially varies along each pole pair.}\label{fig:example_ripple}
    \end{figure}

Hall-based sensing nevertheless suffers from position-dependent inaccuracies due to uneven magnetization, manufacturing tolerances, and sensor misalignments. These imperfections introduce periodic measurement errors, which can lead to degraded control performance and parasitic vibrations~\citep{Pan2015,Xiao2007}, see Figure~\ref*{fig:example_ripple}. Calibration is thus required to eliminate the resulting ripples in the estimated rotor position.

Existing approaches to sensor calibration use external sensors or automated test benches to obtain a ground truth~\citep{Dresscher2019,meer_ifac2023}, 
which effectively corrects measurement errors. Alternatively, filter-based methods~\citep{Xiao2007,Jung2010} successfully suppress Hall-induced vibrations online using feedback. Other methods avoid external sensors by using measurement models~\citep{Du2018,Kim2016} or extended Kalman filters~\citep{Zhao2004}. Still, these methods have their limitations. Reliance on external sensors greatly increases the cost of calibration in a mass-production setting
, even with automated test benches. Moreover, filter-based methods limit control bandwidth by introducing phase lag. Existing methods avoiding external position sensors instead rely on rough position estimates, assume ideal sensor placement or are too computationally demanding for low-cost hardware.  

Although these methods improve measurement accuracy, no procedure relies solely on analog Hall signals and actuator torque commands while avoiding strict assumptions on sensor placement. Therefore, this \manuscript aims to calibrate analog linear Hall sensors through closed-loop experiments and simulation error minimization. No external angle sensor or expensive test setup is needed, making the method suitable for cost-sensitive, large-scale production.

The main contributions are as follows.
\begin{enumerate}[label=C\arabic*]
\item A closed-loop identification and calibration strategy is developed that relies solely on Hall measurements and torque commands while capturing higher-order harmonic distortions in the flux density.
\item Simulation results show that the method accurately estimates the rotor angle without external position information.
\item Experiments demonstrate improved measurement accuracy on an industrial setup.
\end{enumerate}
This \manuscript is structured as follows. Section~\ref*{sec:problem} formalizes the problem. Section~\ref*{sec:method} describes the calibration approach. Sections~\ref*{sec:sim} and~\ref*{sec:exp} demonstrate its effectiveness in simulations and experiments. Finally, Section~\ref*{sec:conclusion} provides conclusions.
 

\section{Problem Description}\label{sec:problem}
This section describes the challenges associated with reconstructing the rotor position of an electric motor using Hall sensor measurements. 

\subsection{Experimental setup: Hall sensors on an electric motor}
Consider an electric motor with linear time-invariant (LTI) torque dynamics given by
\begin{equation}
    y_0(s) = G(s) T(s),
\end{equation}
where $y_0 \in \mathbb{R}$ represents the true rotor position, 
\begin{equation}
T(s) = T_u(s) + T_d(s)
\end{equation}
is the applied torque consisting of a control action $T_u$ and external disturbances $T_d$, and $G(s)$ is a transfer function with Laplace operator $s$. The rotor contains $n_m$ pole pairs that generate a position-dependent magnetic field. Three Hall sensors $h \in \{1, 2, 3\}$ are mounted on the stator, spaced approximately 120$^\circ$ apart in electrical angle. Neglecting dependence on temperature, each sensor measures a voltage $d_h$ assumed proportional to the local magnetic flux density, given by 
\begin{equation}
    d_h(t_k) = g_h(y_0(t_k)) + v_h(t_k),
\end{equation}
where $t_k = T_s k$ with sample time $T_s$ and discrete-time sample number $k$. Here, $g_h(y_0)$ describes the periodic relationship between rotor position $y_0$ and scaled flux density with $y_0=0$ at $t_0$, and $v_h(t_k)$ is zero-mean, independent sensor noise with variance $\sigma_h^2$. The series connection of linear system $G(s)$ and nonlinear functions $g_h(y_0)$ is recognized as a single-input multi-output Wiener system in literature~\citep{Westwick1996}.

\subsection{Computing the rotor position from Hall sensor data}
Estimates $y\approx y_0$ can be reconstructed from the Hall sensor measurements $d_h$ if the mapping \begin{equation}
    \mathbf{g}(y_0) = \begin{bmatrix}
        g_1(y_0) & g_2(y_0) & g_3(y_0)
    \end{bmatrix}^\top
\end{equation} 
has a left inverse. This is the case if and only if $\mathbf{g}(y_0)$ is injective, i.e., any unique flux density vector $\mathbf{d}=\mathbf{g}(y_0)$ must correspond to exactly one rotor position $y_0$. This is not the case on the whole domain $y_0\in\mathbb{R}$: not only is $\mathbf{g}(y_0)$ periodic with mechanical period $2\pi$, it is also periodic with period $\frac{2\pi}{n_m}$ if the pole-pairs are placed axisymmetrically. 

This issue is overcome by including prior information about the specific period that $y_0(t_k)$ is currently in, e.g., by using the previous position estimate \begin{equation}
    \phi\triangleq y(t_{k-1}) 
\end{equation} and assuming a sufficiently small $T_s$. In this case, $\mathbf{g}(y_0)$ is not required to be injective on the whole domain $y_0\in\mathbb{R}$, but only in a domain $\mathcal{Y}_\phi$ smaller than the periodicity of $\mathbf{g}(y_0)$, centered around $\phi$: 
\begin{equation}\label{eq:domain}
    \mathcal{Y}_\phi = \left\{y_0 \mid \phi - \frac{\pi}{n_m} < y_0 < \phi + \frac{\pi}{n_m}\right\}.
\end{equation}
Within this domain, the estimate $y$ of the true position $y_0$ follows from a function $f_\phi$ satisfying
\begin{equation}\label{eq:invert_true}
    f_\phi(\mathbf{g}(y_0)) = y_0, \quad \forall y_0 \in \mathcal{Y}_\phi.
\end{equation}
Thus, $\phi\triangleq y(t_{k-1})$ in~\eqref{eq:invert_true} acts as a history-capturing variable that enables reconstruction of the mechanical rotor position $y_0$ despite periodic flux densities.

Since $\mathbf{g}(y_0)$ is unknown,~\eqref{eq:invert_true} cannot be used for designing the left inverse $f_\phi$. Instead, $f_\phi$ is designed using a model $\hat{\mathbf{g}}(y_0)\approx\mathbf{g}(y_0)$ to satisfy the condition \begin{equation}\label{eq:invert_model}
    f_\phi(\hat{\mathbf{g}}(y_0)) = y_0, \quad \forall y_0 \in \mathcal{Y}_\phi,
\end{equation}
where model mismatch would lead to estimation error $y_0-f_\phi(\hat{\mathbf{g}}(y_0))$. The next section addresses the importance of accurately modeling $\hat{\mathbf{g}}(y_0)\approx\mathbf{g}(y_0)$.

\subsection{Consequences of incorrect reconstruction}\label{sec:misalign}
Imperfect modeling of $\mathbf{g}(y_0)$ leads to periodic errors in the reconstructed rotor position $y$, resulting in ripples that degrade tracking performance and cause vibrations. Assuming Hall signals are purely sinusoidal is inadequate due to manufacturing tolerances, uneven magnetization, and misaligned sensors. These imperfections introduce higher-order harmonics and cause measurement inaccuracy when left unaddressed; see Figure~\ref*{fig:example_ripple}. This shows the need for a model $\hat{\mathbf{g}}(y_0)$ to accurately capture flux density behavior. 

\subsection{Problem definition}
The aim is to obtain accurate rotor position estimates $y \approx y_0$ from Hall sensor measurements $\mathbf{d}$. No external position sensors are available for calibration except for validation purposes, and the solution must be robust to external disturbances and implementable on low-cost embedded hardware. This involves two main tasks:
\begin{enumerate}
    \item Identify an accurate flux density model $\hat{\mathbf{g}}(y_0)$ based on the measurements $\mathbf{d}$ and applied torque $T$.
    \item Design a
n $f_\phi$ satisfying~\eqref{eq:invert_model}.
\end{enumerate}

\section{Self-calibrating Hall sensors}\label{sec:method}
This section describes the developed calibration approach for Hall-based rotor position estimation. Section~\ref{sec:modeling} presents flux density modeling, experiment design, and identification. Section~\ref{sec:reconstruction} details the reconstruction function, and Section~\ref{sec:impl} covers implementation. Algorithm~\ref{algo:hall} summarizes the procedure.

\begin{algorithm}[tb]
    \caption{Data-driven calibration of Hall sensors}\label{algo:hall}
    \begin{algorithmic}[1]
    \Require Controller $C(s)$, BLA $\hat{G}_{\text{BLA}}(q)$, reference $r(t_k)$.
    \State Track $r(t_k)$ in closed-loop using $f_\phi=f_\phi^{\text{init}}$ in~\eqref{eq:fimp}, store $\mathbf{d}(t_k)$ and $y(t_k)$ in $\mathcal{D}$. \hfill(Section~\ref{sec:modeling}) 
    \State Set $\hat{G}_{\text{BLA}}(q)\leftarrow \hat{G}_{\text{BLA}}(q)/\hat{c}$ with~\eqref{eq:c_hat}.\hfill(Section~\ref{sec:impl})\label{step:correct}
    \State Solve~\eqref{eq:problem} to obtain $\hat{\mathbf{g}}_{\theta^\star}$.\hfill(Section~\ref{sec:modeling})\label{step:model}
    \State Create $f_\phi^{\star}$ using~\eqref{eq:f_final}.\hfill(Section~\ref{sec:reconstruction})
    \State \Return Reconstruction function $f_\phi^{\star}$. 
        \end{algorithmic}
    \end{algorithm}   

    \begin{figure}
        \centering
        \includegraphics[width=\linewidth]{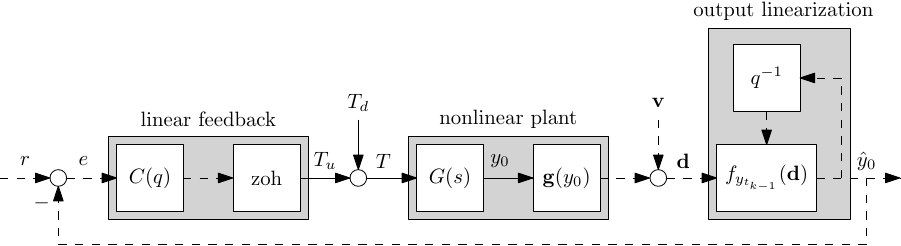}
        \caption{Closed-loop data collection scheme. Rotor position estimates $y\approx y_0$ are reconstructed from flux density signals $\mathbf{d}$ and are used for position feedback control, suppressing external disturbances $T_d$. Solid and dashed lines represent continuous-time and discrete-time signals, respectively.}\label{fig:cl_scheme}
        \end{figure}

\subsection{Modeling the flux density function $\mathbf{g}$}\label{sec:modeling}
The first step involves identifying an accurate model $\hat{\mathbf{g}}(y_0)$ of the flux density function $\mathbf{g}(y_0)$ from measured data. The modeling process consists of three key steps: experiment design, model structure definition, and identification. These steps are described below.

\subsubsection{Experiment design}\label{sec:expdesign}
Data is collected in closed-loop, using a feedback controller $C(s)$ for safety and for mitigation of disturbances $T_d$. 
During data collection, an initial reconstruction function $f_{\phi}^{\text{init}}$ estimates the rotor position to facilitate linear position feedback. This function combines a Clarke transformation and the $\text{atan2}$ function to approximate the rotor position based on the Hall sensor measurements $\mathbf{d} = [d_1, d_2, d_3]^\top$ \citep{Hussain2016}:  
\begin{equation}\label{eq:fimp}\begin{aligned}
    y(t_k) &= f_{y_{t_{k-1}}}^{\text{init}}(\mathbf{d}(t_k)) \\
    f_{y_{t_{k-1}}}^{\text{init}}(\mathbf{d}(t_k)) &:= \frac{1}{n_m} \left(\Gamma\left(\mathrm{atan2}\left( \tilde{d}_2(t_k),\tilde{d}_1(t_k)\right), y_{t_{k-1}}\right)\right),
\end{aligned}
\end{equation}

with $\Gamma:\mathbb{R}\times \mathbb{R}\to \mathbb{R}$ an unwrapping function given by \begin{equation}
\Gamma(y_{t_k},y_{t_{k-1}}) := y_{t_{k-1}}  + \operatorname{mod}\left(y_{t_{k}} - y_{t_{k-1}} +\pi,2\pi\right)-\pi,
\end{equation}
and $\tilde{\mathbf{d}} = \mathbf{C} \mathbf{d}$ with $\mathbf{C}$ the Clarke transformation matrix:
\begin{equation}
    \mathbf{C}=\frac{2}{3}\begin{bmatrix}
        1 & -\frac{1}{2} &  -\frac{1}{2}\\
    0 &\frac{\sqrt{3}}{2}&-\frac{\sqrt{3}}{2}\\
    \frac{1}{2}&\frac{1}{2}&\frac{1}{2}
    \end{bmatrix}.
\end{equation}
This initial $f_{\phi}^{\text{init}}$ satisfies~\eqref{eq:invert_true} if $\mathbf{g}(y_0)$ consists solely of three pure sinusoids shifted by $120^\circ$, without higher-order harmonics. In practice, however, sensor misalignments and uneven magnetization give rise to harmonic distortions that make~\eqref{eq:fimp} only an approximation $y(t_k)\approx y_0(t_k)$.

Despite these inaccuracies, the approximate reconstruction is sufficient to enable closed-loop control, as shown in Figure~\ref{fig:cl_scheme}. The feedback controller $C$ suppresses external disturbances $T_d$ and ensures that the approximated rotor position tracks a ramp reference. Perfect reference tracking is not achieved because of the higher harmonics in $\mathbf{g}$, but this is not required for identification; the feedback controller need only suppress $T_d$ in $T$. Section~\ref{sec:impl} addresses the reference and controller design. During experiments, the $N$ samples of $\mathbf{d}(t_k)$ and $y(t_k)$ are stored in the dataset $\mathcal{D}$ for use in identification.

\subsubsection{Model structure}
The flux density function $\mathbf{g}(y_0)$ is parametrized linear in the parameters for simplicity: \begin{equation}
    \hat{\mathbf{g}}_{\boldsymbol{\theta}}^\top(y_0) = \boldsymbol{\psi}(y_0) \boldsymbol{\theta},
\end{equation}
where $\boldsymbol{\theta}\in\mathbb{R}^{n_\theta}$ are model parameters, and
\begin{equation}\label{eq:problem}
    \boldsymbol{\psi}(y_0) = \mathbf{I}_3 \otimes \boldsymbol{\beta}(y_0),
\end{equation}
with $\otimes$ the Kronecker product and $\boldsymbol{\beta}: \mathbb{R}\to \mathbb{R}^{1\times m}$ a periodic basis function, such as a Fourier basis or a periodic kernel function~\citep{Rasmussen2004}. The order $m$ can be chosen by analyzing the harmonic content of data $\mathbf{d}(t_k)$.

\subsubsection{Identification}
The parameters $\boldsymbol{\theta}$ are identified by solving a simulation error minimization (SEM) problem. The cost function $J(\boldsymbol{\theta})$ is defined as the squared 2-norm of the difference between the measured flux density $\mathbf{d}\in\mathcal{D}$ and a simulated flux density ${\mathbf{d}}_{\boldsymbol{\theta}}^{\text{sim}}$:
\begin{equation}\begin{aligned}\label{eq:problem}
    \min_{\boldsymbol{\theta}} & \ J(\boldsymbol{\theta}), \\
     J(\boldsymbol{\theta}) = & \sum_{k=1}^N  \left(\mathbf{d}(t_k) - {\mathbf{d}}_{\boldsymbol{\theta}}^{\text{sim}}(t_k)\right)^\top \left(\mathbf{d}(t_k) - {\mathbf{d}}_{\boldsymbol{\theta}}^{\text{sim}}(t_k)\right) .
\end{aligned}\end{equation}
Here, ${\mathbf{d}}_{\boldsymbol{\theta}}^{\text{sim}}(t_k)$ is computed using the state and output equations of the closed-loop system, incorporating a Best Linear Approximation (BLA) $\hat{G}_{\text{BLA}}(q)\approx G(q)$, detailed in Section~\ref{sec:impl}, and the model $\hat{\mathbf{g}}_{\boldsymbol{\theta}}(y_0)$. These equations are:
\begin{align}
    {\mathbf{d}}_{\boldsymbol{\theta}}^{\text{sim}}(t_k) &= \hat{\mathbf{g}}_{\boldsymbol{\theta}}(y_0^{\text{sim}}(t_k)),\notag\\
    y_{0,\boldsymbol{\theta}}^{\text{sim}}(t_k) &= \mathbf{C}_{G_{\text{BLA}}} \mathbf{x}_{G_{\text{BLA}},\boldsymbol{\theta}}(t_k),\notag\\
    \mathbf{x}_{G_{\text{BLA}},\boldsymbol{\theta}}(t_{k+1}) &= \mathbf{A}_{G_{\text{BLA}}} \mathbf{x}_{G_{\text{BLA}},\boldsymbol{\theta}}(t_k) + \mathbf{B}_{G_{\text{BLA}},\boldsymbol{\theta}} T_{\boldsymbol{\theta}}^{\text{sim}}(t_k),\notag\\
    T_{\boldsymbol{\theta}}^{\text{sim}}(t_k) &= \mathbf{C}_C \mathbf{x}_{C,\boldsymbol{\theta}}(t_k) + \mathbf{D}_C e_{\boldsymbol{\theta}}^{\text{sim}}(t_k),\notag\\
    \mathbf{x}_{C,\boldsymbol{\theta}}(t_{k+1}) &= \mathbf{A}_C \mathbf{x}_{C,\boldsymbol{\theta}}(t_k) + \mathbf{B}_C e_{\boldsymbol{\theta}}^{\text{sim}}(t_k),\label{eq:state_eq}\\
    e_{\boldsymbol{\theta}}^{\text{sim}}(t_k) &= r(t_k)-y_{\boldsymbol{\theta}}^{\text{sim}}(t_k),\notag\\
    y_{\boldsymbol{\theta}}^{\text{sim}}(t_k) &= f_{y_{\boldsymbol{\theta}}^{\text{sim}}(t_{k-1})}^{\text{init}}({\mathbf{d}}_{\boldsymbol{\theta}}^{\text{sim}}(t_k)),\quad \forall k\in\{1,\ldots,N\}\notag
\end{align}
with zero initial conditions. Here, $\mathbf{A}_{G_{\text{BLA}}}$, $\mathbf{B}_{G_{\text{BLA}}}$, and $\mathbf{C}_{G_{\text{BLA}}}$ represent the state and output matrices of the BLA, while $\mathbf{A}_C$, $\mathbf{B}_C$, $\mathbf{C}_C$, and $\mathbf{D}_C$ describe the discrete-time controller. Problem~\eqref{eq:problem} is solved using an interior-point method with approximate gradients, starting at an initial estimate $\boldsymbol{\theta}_0$ corresponding to pure sinusoids, i.e., the inverse of $f^{\text{init}}$. Since the cost is non-convex, $\boldsymbol{\theta}^\star$ corresponds to a local minimum.

\subsection{Designing a reconstruction function $f_\phi^\star$}\label{sec:reconstruction}
With the flux density model $\hat{\mathbf{g}}(y_0)$ available, a reconstruction function $f_\phi^\star$ is designed to estimate the rotor position while compensating for inaccuracies in the initial reconstruction function $f_\phi^{\text{init}}$. First, a lookup table is defined on a grid of $M$ equidistant points $y_{0,i}$ within the interval $[0,2\pi)$. Using the model $\hat{\mathbf{g}}(y_0)$, the corresponding outputs of the initial reconstruction function $f_\phi^{\text{init}}$ are computed:
\begin{equation}\label{eq:recon_biject}
\hat{y}_i^{\text{}} = f_{y_{0,i}}^{\text{init}}(\hat{\mathbf{g}}_{\boldsymbol{\theta}}(y_{0,i})),\quad i\in\{1,\ldots,M\}.
\end{equation}
The additive measurement error caused by $f_\phi^{\text{init}}$ on this grid is then estimated as \begin{equation}
    \hat{\eta}_i^{\text{LUT}} := y_{0,i} - \hat{y}_i^{\text{}},\quad i\in\{1,\ldots,M\}.
\end{equation}
Next, a piecewise-linear correction function $\hat{\eta}^{\text{LUT}}(\hat{y})$ is defined to interpolate between the points $(\hat{y}_i, \eta_i^{\text{LUT}})$. For a given $\hat{y}$, the index $i$ is determined such that:
\begin{equation}
    i = \operatorname{argmin}_i \left\{\hat{y} \in [\hat{y}_i, \hat{y}_{i+1})\right\},
\end{equation}
where the interpolation wraps around at the boundaries, i.e., $\hat{y}_{M+1} = \hat{y}_1$ and $\eta_{M+1}^{\text{LUT}} = \eta_1^{\text{LUT}}$. The piecewise-linear interpolation is then computed as:
\begin{equation}
    \eta^{\text{LUT}}(\hat{y}) = \hat{\eta}_i^{\text{LUT}} + \frac{\hat{y} - \hat{y}_i}{\hat{y}_{i+1} - \hat{y}_i} \left( \hat{\eta}_{i+1}^{\text{LUT}} - \hat{\eta}_i^{\text{LUT}} \right).
\end{equation}
Finally, the reconstruction function $f_\phi^\star$ is defined to compensate for the additive measurement error, yielding:
\begin{equation}\label{eq:f_final}
    f_\phi^{\star}(\mathbf{d}) = f_\phi^{\text{init}}(\mathbf{d}) + {\eta}^{\text{LUT}}\left(f_\phi^{\text{init}}(\mathbf{d})\right).
\end{equation}
This adjustment corrects the periodic inaccuracies in $f_\phi^{\text{init}}$ in a computationally lightweight manner, improving the accuracy of rotor position estimation. 
Note that this simple approach requires~\eqref{eq:recon_biject} to be bijective. If it is not, a different approach to designing $f_\phi^\star$ is required that avoids $f_\phi^{\text{init}}$ altogether.
The next section discusses relevant implementation aspects.

\subsection{Implementation aspects}\label{sec:impl}
Several practical considerations are important for implementing the developed method, as detailed next.

\subsubsection{Control and reference design}
The closed-loop data collection in Section~\ref*{sec:expdesign} uses a feedback controller $C(s)$ to suppress external disturbances $T_d$ on total torque $T$. To mitigate these disturbances, the sensitivity~\citep{franklinFeedback} must be low in magnitude. With the nonlinear function $\mathbf{g}$ approximately linearized by $f_\phi^{\text{init}}$ in Figure~\ref{fig:cl_scheme}, the sensitivity is given by  
\begin{equation}
S(s) = \frac{T(s)}{T_d(s)} \;\approx\; \frac{1}{1 + G(s)\,C(s)}.
\end{equation}
Including an integrator in $C(s)$ ensures that the sensitivity is low in magnitude at low frequencies and effectively suppresses slowly varying disturbances. A slow ramp reference from 0 to $2\pi n$\ rad with $n\in\mathbb{R}_{\geq 1}$ places any position-dependent disturbances in this low-frequency range where they are well attenuated. 

\subsubsection{Obtaining a Best Linear Approximation}\label{sec:BLA}
The BLA $\hat{G}_{\text{BLA}}(q)$ required for Step~\ref*{step:model} in Algorithm~\ref*{algo:hall} is identified up to an unknown constant in closed-loop using the approach described in~\citet[Section 3.8]{Pintelon2012}, averaging over multiple realizations of random-phase multisine reference signals. This yields \begin{equation}
    \hat{G}_{\text{BLA}}(q)=c G(q),
\end{equation} with $c\in\mathbb{R}$. To correct for this mismatch, $\hat{c}\approx c$ is estimated from data $\mathcal{D}$ by minimizing 
\begin{equation}\label{eq:c_hat}\begin{aligned}
\hat{c} =& \arg\min_c\ \left(y_{\boldsymbol{\theta}_0,c}^{\text{sim}}(t_{p_{\text{sim}}})-y^{\text{dat}}(t_{p_{\text{dat}}})\right)^2,
\end{aligned}
\end{equation}
where $y_{\boldsymbol{\theta}_0,c}^{\text{sim}}(t_{p_{\text{sim}}})$ follows from~\eqref{eq:state_eq} using a scaled BLA $\tilde{\mathbf{B}}_{G_{\text{BLA}}}=c\mathbf{B}_{G_{\text{BLA}}}$, and $y^{\text{dat}}\in\mathcal{D}$. Moreover, $t_p$ is the time instance of the last full rotation:
\begin{equation}
p_\iota := \arg\max_{p_\iota} \left\{y^{\iota}(t_{p_\iota}) = 2\pi n \mid n\in\mathbb{N}, t_{p_\iota}\leq t_N\right\},
\end{equation}
for both $\iota\in\{\text{sim},\text{dat}\}$. The estimate $\hat{c}\approx c$ in~\eqref{eq:c_hat} relies on $\mathbf{g}(y_0)$ being periodic with known period $2\pi$. Indeed, the number of full rotations of the rotor is independent of the shape of the measurement errors, so any mismatch between simulation~\eqref{eq:state_eq} and the data must be attributed to incorrect scaling of the BLA. Once $\hat{c}\approx c$ is estimated, it is used to compensate the BLA in Step~\ref*{step:correct} of Algorithm~\ref*{algo:hall}, before the nonlinear identification step.

\section{Simulation results}\label{sec:sim}
This section demonstrates the performance of the developed calibration approach on a simulation example.

\subsection{Simulation setup}
Consider an example motor with \(n_m=11\) permanent magnets and linear dynamics
\begin{equation}
G(s) := \frac{y_0(s)}{T(s)} = \frac{1.663 \cdot 10^5}{s^3 + 632.6\,s^2 + 2702\,s} \exp \bigl(1.2 \cdot 10^{-4} s\bigr).
\end{equation} 
Moreover, \(\mathbf{g}(y_0)\) is parametrized by a Fourier basis \(\boldsymbol{\beta}(y_0) \in \mathbb{R}^{1 \times (2n_h+1)}\). The first element is \(\beta_1(y_0) = 1\), and
\begin{equation} \label{eq:basis}
    \begin{aligned}
    \beta_{1+i}(y_0) &= \sin\bigl(y_0\, h_{\lceil i/2 \rceil}\bigr),\\
    \beta_{2+i}(y_0) &= \cos\bigl(y_0\, h_{\lceil i/2 \rceil}\bigr),\\
    &\forall i \in \{\,i \in \mathbb{N}\mid i \text{ odd},\, 1 \le i \le 2n_h-1\},
    \end{aligned}
    \end{equation}
    with harmonics \(\mathbf{h} = [\,1,\ldots,11\,]^\top\). A parameter vector \(\boldsymbol{\theta}\) is chosen so that each permanent magnet has a slightly different flux density profile, as Figure~\ref{fig:sim_g} shows it is not quite periodic with period \(2\pi/n_m\). The flux density signals are sampled at \(F_s = 4000\) Hz, each with noise variance \(\sigma_h^2 = 7.5 \cdot 10^{-6}\,\mathrm{V}\). A stabilizing controller is given by  
    \begin{equation}\label{eq:controller}
    C(q) = \frac{2.94 q^3 - 3.29q^2-2.10q+2.45}{q^4-3.45q^3+4.52q^2-2.68q+0.61}
    \end{equation}
and is used in the control scheme in Figure~\ref{fig:cl_scheme}.

\subsection{Approach}
First, a BLA \(\hat{G}_{\mathrm{BLA}}(s)\) is measured in closed-loop following the procedure in Section~\ref{sec:BLA}. Algorithm~\ref{algo:hall} is then applied with model structure~\eqref{eq:basis}. During data collection, a reference \(r(t_k)\) increases linearly from 0 to 13\,rad in 26\,s. Problem~\eqref{eq:problem} is solved in two hours on a standard desktop computer, and~\eqref{eq:recon_biject} is verified to be bijective. 

\subsection{Results}
\begin{figure}
    \centering
    \includegraphics[width=\mylinewidth]{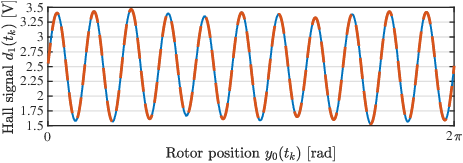}
    \caption{Simulation example. The first Hall signal $d_1(t_k)$ (\protect\blueline) is approximately periodic with the magnet pitch, with slight variations across magnets. The estimates $\hat{d}_1=\hat{{g}}_{1,\boldsymbol{\theta}^\star}(\mathbf{d}(t_k))$ (\protect\reddash) from the model in Step~\ref{step:model} of Algorithm~\ref*{algo:hall} closely match the true function.}\label{fig:sim_g}
    \end{figure}

\begin{figure}
        \centering
        \includegraphics[width=\mylinewidth]{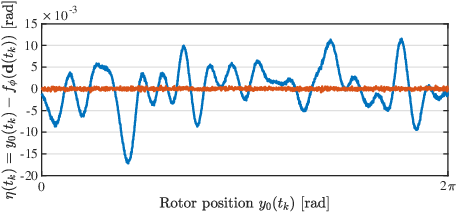}
        \caption{Measurement error in the simulation. Using the initial $f_\phi^{\text{init}}(\mathbf{d})$ results in a large measurement error (\protect\blueline). Using $f_\phi^{\star}(\mathbf{d})$ from Algorithm~\ref*{algo:hall} reduces the measurement error down to the noise floor (\protect\redline).}\label{fig:sim_eta}
        \end{figure}

Figure~\ref{fig:sim_g} illustrates one of the simulated flux density signals \(d_1(t_k)\), together with the estimates \(\hat{d}_1(t_k) = \hat{g}_{1,\boldsymbol{\theta}^\star}(y_0^{\mathrm{sim}}(t_k))\). The model \(\hat{\mathbf{g}}_{\boldsymbol{\theta}^\star}\) accurately captures the slight flux density variations across the magnets. Figure~\ref{fig:sim_eta} depicts the estimation error in the rotor angle when using the initial reconstruction \(f_\phi^{\mathrm{init}}\) versus the final reconstruction \(f_\phi^\star\). The initial reconstruction exhibits a clear periodic error due to unmodeled higher-order harmonics. The final reconstruction \(f_\phi^\star\) corrects these structural errors and reduces the error to the sensor noise level. Note that the true rotor position \(y_0\) is only used here for validation; it is not part of the calibration procedure.

These results show that the developed method accurately calibrates Hall sensors without relying on an external reference encoder. As shown next, measurement accuracy is also improved on an industrial setup.

\section{Experimental results}\label{sec:exp}
This section validates the approach experimentally.
\subsection{Experimental setup}
A confidential setup from Sioux Technologies B.V. with a Brushless Direct Current (BLDC) motor is used for experimental validation. The setup follows Figure~\ref*{fig:hall_scheme}, with a rotor of \(n_m = 11\) pole pairs and an external encoder for validation only. Three Hall sensors, spaced approximately 120 electrical degrees apart, are sampled at \(F_s = 4000\)\,Hz. The same feedback controller as in~\eqref{eq:controller} is used.

\begin{figure}
    \centering
    \includegraphics[width=\mylinewidth]{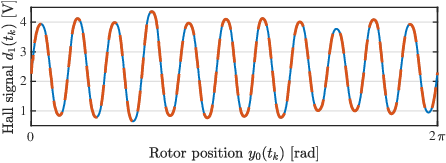}
    \caption{Experimental data. The measured Hall signal ${d}_1(t_k)$ (\protect\blueline) repeats roughly with each magnet but shows slight variations. The identified model $\hat{d}_1(t_k) = \hat{{g}}_{1,\boldsymbol{\theta}^\star}(y_0^{\mathrm{sim}}(t_k))$ (\protect\reddash) accurately estimates the flux densities without relying on $y_0(t_k)$.}\label{fig:exp_g}
    \end{figure}

    \begin{figure}
        \centering
        \includegraphics[width=\mylinewidth]{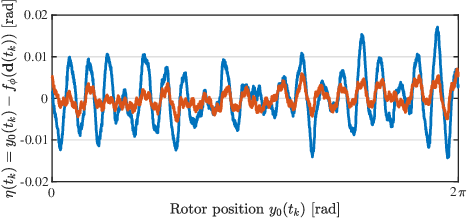}
        \caption{Measurement error in the experiments, with the external encoder used for validation only. The initial $f_\phi^{\text{init}}(\mathbf{d})$ produces a large error (\protect\blueline). After Algorithm~\ref*{algo:hall}, $f_\phi^{\star}(\mathbf{d})$ achieves a significant reduction (\protect\redline).}
        \label{fig:exp_eta} 
        \end{figure}

        \begin{figure}
            \centering
            \includegraphics[width=\mylinewidth]{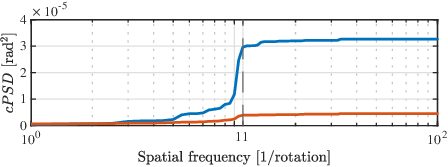}
            \caption{Cumulative power spectral density of the measurement error. The initial $f_\phi^{\text{init}}(\mathbf{d})$ (\protect\blueline) shows clear periodic content with the spatial frequency $n_m=11$ corresponding to the magnet count. The final $f_\phi^{\star}(\mathbf{d})$ (\protect\redline) corrects for these errors.}
            \label{fig:exp_spectrum}
            \end{figure}

\subsection{Approach}

As before, a BLA \(\hat{G}_{\mathrm{BLA}}(s)\) is identified using the closed-loop approach in Section~\ref{sec:BLA}. The flux density model \(\hat{\mathbf{g}}_{\boldsymbol{\theta}}\) is then expressed through a kernel-based basis function 
\(\boldsymbol{\beta}(y_0) : \mathbb{R} \to \mathbb{R}^{1 \times m}\), 
where \(m = 400\). A grid of \(m\) points \(y_{0,j}\) is defined equidistantly in \([\,0, 2\pi)\). The kernel is defined by  
\begin{equation}
    \boldsymbol{\beta}_j(y_0) = k(y_0,y_{0,j}), 
\quad 
    k(y,y') = \sigma_f^2 \exp\!\Bigl(-\tfrac{1}{2\ell^2}\|\mathbf{x}-\mathbf{x}'\|^2\Bigr),
\end{equation}
where \(\mathbf{x} = [\sin(y),\,\cos(y)]^\top\). Hyperparameters \(\sigma_f\) and \(\ell\) are selected by including them as design variables in~\eqref{eq:problem}. The reference \(r(t_k)\) is a ramp from 0 to 20\,rad over 40\,s. Problem~\eqref{eq:problem} is solved in ten hours on a standard desktop computer, and~\eqref{eq:recon_biject} is verified to be bijective.

\subsection{Results}
Figure~\ref{fig:exp_g} shows an example of the measured Hall signal \(d_1(t_k)\) and its estimate \(\hat{d}_1(t_k) = \hat{g}_{1,\boldsymbol{\theta}^\star}(y_0^{\mathrm{sim}}(t_k))\), where the identified model captures magnet variations. Figure~\ref{fig:exp_eta} presents the measurement error using the external encoder for validation. The initial reconstruction \(f_\phi^{\mathrm{init}}\) results in an RMS error of 5.7\,mrad, while the final reconstruction \(f_\phi^\star\) compensates for higher-order harmonics, reducing it to 2.2\,mrad. The peak \(\|\eta\|_\infty\) is reduced by a factor of 2.5. 

Figure~\ref{fig:exp_spectrum} shows the cumulative power spectral density of the measurement error
. Much of the frequency content aligns with the magnet pitch, which the corrected reconstruction significantly suppresses. These results confirm that the calibration method improves rotor position estimation on an industrial setup, achieving a factor of 2.6 improvement in RMS accuracy and a factor of 2.5 in peak error without requiring an external reference encoder.

\subsection{Discussion} 
The residual errors in Figure~\ref{fig:exp_eta} are presumably caused by nonlinear dynamics that are periodic in $y_0$ with period $2\pi$, such as cogging, affecting $\mathbf{d}$ and indistinguishable from the contribution of $\mathbf{g}$. A potential solution might be to repeat the data collection process for different angular placements of the motor coils, averaging out this effect. This would require a modular design and involves further research.   

Furthermore, the developed two-step approach could potentially be simplified to directly construct $f_\phi$ from data, avoiding modeling the of $\mathbf{g}$. The current two-step approach is motivated by the expectation that nonlinear identification through simulation-error minimization is more robust to measurement noise on $\mathbf{d}$, yet a more thorough analysis for this choice is desirable.

\section{Conclusion}\label{sec:conclusion}
The developed method improves measurement accuracy of Hall sensors without using external encoders, improving positioning performance and reducing vibrations cost-effectively for mass production. The simulation error minimization accurately estimates flux density functions, and the resulting compensation function reduces measurement error by a factor of 2.6 on an industrial setup. These findings eliminate the need for expensive test benches and enable low-cost position measurements. Future work will focus on reducing offline computation time through multiple shooting and lower-dimensional model structures, and an extension to Hammerstein systems.

\begin{ack}
    The authors thank Sioux Technologies B.V. for the developments that led to these results and for their support in carrying out the experiments reported in this \manuscript.
    \end{ack}

{\footnotesize
\bibliography{library.bib}
}

\end{document}